\documentclass[openacc]{rstransa}

\newtheorem{theorem}{\bf Theorem}[section]

\usepackage{amsmath,amssymb,amsfonts,amsthm,graphicx,mathrsfs,setspace,verbatim,times,helvet,courier,bm,url,dcolumn}

\usepackage[english]{babel}

\def\be{\begin{equation}}
\def\ee{\end{equation}}
\def\bea{\begin{eqnarray}}
\def\eea{\end{eqnarray}}
\def\bean{\begin{eqnarray*}}
\def\eean{\end{eqnarray*}}

\def\scri{\mathscr{J}}

\begin{document}

\title{A critical appraisal of the singularity theorems}

\author{
Jos\'e M. M. Senovilla$^{1}$
}

\address{$^{1}$Physics Department, University of the Basque Country UPV/EHU, Apartado 48080, Bilbao, Spain}

\subject{Physics, Gravity, Relativity}


\corres{Jos\'e M. M. Senovilla\\
\email{josemm.senovilla@ehu.eus}}

\begin{abstract}
The 2020 Nobel prize in Physics has revived the interest in the singularity theorems and, in particular, in the Penrose theorem published in 1965. In this short paper I briefly review the main ideas behind the theorems and then proceed to an evaluation of their hypotheses and implications. I will try to dispel some common misconceptions about the theorems and their conclusions, as well as to convey some of their rarely mentioned consequences. In particular, a discussion of spacetime extensions in relation to the theorems is provided. The nature of the singularity inside black holes is also analyzed.
\end{abstract}


\begin{fmtext}
\section{Introduction}
Penrose's singularity theorem \cite{P,SMilestone} won (half of) the 2020 Nobel prize in Physics! It was a \underline{deserved} award. The theorem is a beautiful piece of mathematical physics,
contained novel ingredients along with fruitful ideas, prompted multiple developments in theoretical relativity, and involved stunning physical consequences.

In particular, the fundamental notion of {\em trapped spheres} was introduced for the first time in \cite{P}. This eventually became a key idea in black-hole physics \cite{H4,P4,HayBook}, numerical relativity \cite{BSh,Thor,JVG}, mathematical relativity \cite{JVG,Kri}, cosmology, gravity analogs \cite{BLV}, etc. Its influence is endless, and its prolific range of applications keeps growing \cite{Bengtsson,S2,HayBook}.

As argued elsewhere \cite{S5,SMilestone}, singularity theorems constitute the first post-Einsteinian content of relativity. What had happened before the theorem (deflection of light, universal expansion, Cauchy's problem, ADM formulation, Kerr's solution, Wheeler-de Witt equation, ...) as well as most of what came next (cosmic background radiation, gravitational lenses, radiation in binary systems, direct observation of gravitational waves, etc) had been explicitly predicted, known, anticipated, or foreseen in one way or another by Einstein. In contrast, the singularity theorems, the ideas sustaining them, and their consequences were (surely) not imagined by the founder of General Relativity (GR). In 1965 GR was emancipated and became a mature theory full of vitality and wonders.
\end{fmtext}


\maketitle

\section{The Penrose singularity theorem}
Penrose's singularity theorem reads, in modern terms, as follows
\begin{theorem}[Penrose singularity theorem]\label{genpen}
In a spacetime of sufficient differentiability, if
\begin{itemize}
\item the null convergence condition holds
\item there is a non-compact Cauchy hypersurface $\Sigma$
\item  and a closed future-trapped surface,
\end{itemize}
then there are future-incomplete null geodesics.
\end{theorem}
There are two important novelties here:
\begin{enumerate}
\item Use of geodesic incompleteness as characterization of spacetime failure
\item Concept of closed trapped surface
\end{enumerate}
Before entering into the details of these two fundamental advances, a brief explanation about the rest of assumptions is in order. A Cauchy hypersurface is a spacelike slice amenable to sustain initial data that determine the full spacetime completely. In simpler words, it is a spacelike hypersurface such that every inextensible timelike or null curve crosses it exactly once. Spacetimes with Cauchy hypersurfaces are called {\em globally hyperbolic} \cite{HE,P5,Wald}. Assuming that $\Sigma$ is non compact amounts to saying that the spacetime is open, that is to say, space is not finite.

The {\em convergence condition} arises as a requirement from the Raychaudhuri equation \cite{Ray,Ray2,K}:
$$
v^{\nu}\nabla_{\nu}(\nabla_{\mu}v^{\mu})+\nabla_{\mu}v^{\nu}\nabla_{\nu}v^{\mu}-
\nabla_{\mu}(v^{\nu}\nabla_{\nu}v^{\mu})+R_{\rho\nu}v^{\rho}v^{\nu}=0.
$$
Here $v^\mu$ is a (null or unit timelike) vector field tangent to a pencil of geodesics, $R_{\mu\nu}$ the Ricci tensor and $\nabla_\mu$ the Levi-Civita covariant derivative. If the geodesics emanate either from a point or orthogonal to a (hyper)surface
then $v^{\nu}\nabla_{\nu}v^{\mu}=0$ and $\nabla_\mu v_\nu =\nabla_{(\mu} v_{\nu)}$ so that the equation becomes 
$$
v^{\nu}\nabla_{\nu}(\nabla_{\mu}v^{\mu})=-\nabla_{\mu}v_{\nu}\nabla^{\mu}v^{\nu}-
R_{\rho\nu}v^{\rho}v^{\nu}
$$
and one knows that first term of the righthand side is non negative. If one also assumes that the second term is non-negative, it is easily proven that if
$\nabla_\mu v^\mu |_p<0$ at any point $p$ then $\nabla_\mu v^\mu \rightarrow -\infty$ in 
finite proper time (or affine parameter for null $v^\mu$). In general, these are called {\em caustics, or focal points}. Geodesics stop maximizing the interval if they encounter one of these focal points. This is called the focusing effect of gravity. Of course, one needs the geometric condition
\be\label{ConCon}
R_{\rho\nu}v^{\rho}v^{\nu}\geq 0 
\ee
usually called the null or timelike convergence condition if $v^\mu$ is null or timelike, respectively.

In GR, one can relate the Ricci tensor to the energy-momentum tensor $T_{\mu\nu}$ via 
Einstein's field equations
\be\label{eq:EFE}
R_{\mu\nu}-\frac{1}{2}R\,  g_{\mu\nu}+\Lambda g_{\mu\nu}=\frac{8\pi 
G}{c^4}T_{\mu\nu}
\ee
where $R$ is the scalar curvature, $G$ is Newton's gravitational 
constant, $c$ is the speed of light in vacuum and $\Lambda$ the 
cosmological constant. 
Thereby, the convergence condition can be 
rewritten in terms of physical quantities. This is why 
$R_{\rho\nu}v^{\rho}v^{\nu}\geq 0$, when valid for all time-like $v^{\mu}$, is called the 
strong energy condition. 
One should bear in mind, however, that this is a condition on the Ricci tensor (a geometrical 
object) and thus it can be used in general geometric theories of gravity.
Observe that, in GR, $\Lambda$ makes no difference for \underline{null} geodesics.

\subsection{Incompleteness}
Spacetime breaks down at singularities, hence singularities are not in spacetime. This is one of the great difficulties in trying to define what is a singularity in GR \cite{Ge2}.
Penrose's idea was to use physical curves that do belong to the spacetime, because curves are good  pointers. If such curves cannot be continued in a regular manner they are pointing towards a problem: the singularity. In his theorem Penrose used hypothetical light rays --null geodesics-- that reach a sudden end, so that one may interpret that they are actually ``aiming at singularities”.

Resorting to geodesic incompleteness was a clever innovation that avoided the multiple difficulties for a rigorous
definition of singularity in GR. Hence, it eventually became (under the influence of Hawking \cite{HE} and 
Geroch \cite{Ge2}) the standard characterization of singularities proven by the singularity theorems.
Usually, only time-like or null curves are used, but in principle also incomplete space-like curves define singularities. 
Moreover, the curves should not need to be geodesic, and as a matter 
of fact there are known examples \cite{Ge2} of geodesically complete space-times 
with incomplete time-like curves of everywhere bounded acceleration. 
It must be remarked, however, that all singularity theorems to date prove merely the existence of geodesic incompleteness, which of course is sufficient proof of incompleteness.

\subsection{Trapped spheres}
The concept of closed trapped surface, in short trapped sphere if the surface has $\mathbb{S}^2$ topology, was a genius idea that characterizes the “point of no return” in stellar gravitational collapse and actually changed the path of GR for ever more.

In GR gravity is described by the geometry of spacetime. Hence, in dynamical situations geometrical quantities such as
area or volume do evolve with time. In very extreme cases, the area of spheroids may decrease to the future {\em no matter how they choose to evolve} ---keeping causality. In this sense, they can be considered ``trapped'', or rather trapping, as everything they contain is doomed to be surrounded by spheroids of less and less area... until a catastrophe may occur. Penrose proved that, when trapped spheres form, incompleteness of spacetime generically develops.

From the geometrical point of view, trapped spheres can be defined by using classical geometrical quantities. Let $\zeta$ be any spacelike submanifold of any dimension in spacetime, and denote by $A_\zeta$ its ``area, volume, etc'' depending on its dimension. Choose an arbitrary vector field  $\xi^\mu$ and deform $\zeta$ along its flow. The initial variation $\delta_\xi A_\zeta$ of $A_\zeta$ due to this deformation is \cite{Kri,MS,S2,SMilestone} 
$$
\delta_\xi A_\zeta = \int_\zeta \left({\rm div} \xi^T + H^\mu \xi_\mu \right)
$$
where $H^\mu$ is the mean curvature vector of $\zeta$, that is, the trace of its second fundamental form (or shape tensor) and $\xi^T$ is the component of $\xi^\mu$ tangent to $\zeta$ (div is the divergence operator in $\zeta$). Recall that $H^\mu$ is orthogonal to $\zeta$. 
Thus, if $\zeta$ is compact the first term in the variation integrates out and one simply has
$$
\delta_\xi A_\zeta = \int_\zeta H^\mu \xi_\mu .
$$
This is a classical result in Riemannian geometry from where minimal submanifolds are characterized by $ H^\mu =0$, and this is the only distinguished case for positive-definite metrics.
However, in Lorentzian geometry if, say, $H^\mu$ is future timelike on $\zeta$ then {\em the variation of $A_\zeta$ along \underline{any future-directed} vector field $\xi^\mu$ is strictly negative}, because then $H^\mu \xi_\mu <0$.
These are precisely the trapped submanifolds: those having $H^\mu$ future timelike. Analogously for past-trapped submanifolds replacing future by past.
As a revealing observation, note that stationary spacetimes cannot have {\em compact} trapped submanifolds \cite{MS} because, if $\xi^\mu$ is chosen to be the timelike Killing vector, $\delta_\xi A_\zeta =0$ ergo $H^\mu$ cannot be future (or past) pointing everywhere on $\zeta$ ---for, in that case, $H^\mu \xi_\mu$ would have a sign.

For each vector field $n^\mu$ normal to $\zeta$, $\theta_n :=H^\mu n_\mu$ is called {\em expansion along $n^\mu$}. Hence, for a future-trapped $\zeta$ \underline{all possible future} expansions $\theta_n$ are negative. This was the original definition by Penrose \cite{P}, where trapped surfaces were characterized by having negative expansions along the null normals.

Observe that this geometric definition of trapped submanifold is independent of the dimension of $\zeta$. One can actually generalize Penrose's theorem to arbitrary spacetime dimension that contain a compact trapped submanifold of any dimension \cite{GaS}.

\subsection{Trapped spheres are stable}
The notion of trapped submanifold is independent of coordinates, bases, existence of symmetries or preferred surfaces, etc.
A decisive point is that its definition is given by {\em inequalities} ($H^\mu H_\mu<0$, or equivalently negative expansions) and therefore trapped submanifolds are {\em stable}: small perturbations of the spacetime will not remove them. 

This is a key point because in 1939 Oppenheimer and Snyder (OS) \cite{OS,OV} proved that spherical collapse of self-gravitating dust (fluid with no pressure) was unstoppable leading to what we call today a {\em black hole}: a region of spacetime fully disconnected from its outside by a one-directional null hypersurface called the {\em event horizon}, see \cite{HE,Wald,H4} for details.. Still, there were some doubts up to what extent the assumptions of spherical symmetry and absence of pressure were determinant for the formation of the black hole. Penrose noticed that the very symmetrical OS model contained trapped spheres. Now, the Einstein-Euler field equations describing a perfect fluid in GR are a system of hyperbolic PDEs, and thus they possess the property of continuous dependence of the solution on the initial conditions. 
The initial conditions of the OS collapse leads to a trapped sphere within a finite time,  hence initial conditions which are sufficiently close to those of OS will also lead to the formation of closed trapped surfaces within the same time interval, {\em regardless of symmetries or of the existence of pressure}. 

\section{Singularity theorems}
S.W. Hawking quickly understood that trapped spheres exist in the standard Fridman-Lema\^\i tre-Robertson-Walker (FLRW) models. Consider the FLRW metric (with flat slices for simplicity)
$$
ds^2 = -c^2 dt^2+ a^2(t) \left(d\chi^2 +\chi^2 d\Omega^2\right)
$$
where $a(t)$ is the scale factor and $d\Omega^2$ the standard metric of a round sphere of unit radius.
Then, round spheres with constant values of $t$ and $\chi$ have the following mean curvature vector (a dot represents derivative with respect to $ct$)
$$
\vec H =\frac{1}{a}\left( -\frac{\dot a}{c} \partial_t + \frac{1}{\chi} \partial_\chi \right)
$$
which clearly is timelike if and only if 
$$
\dot a^2 > 1/\chi^2
$$
 and past or future if $\dot a >0$ or $\dot a <0$ repectively. 
{\em Now}, that is to say, at the present time in the Universe, this can be written 
$$ 
H_0> c/D
$$ 
where $H_0$ is the Hubble constant and $D$ the proper distance to a sphere centred at us.
Thus, for $D$ big enough, such spheres are past trapped if $\dot a >0$.

Hawking and Geroch then proved several theorems of cosmological application \cite{HE}. As an example, the simplest singularity theorem is \cite{HE}
\begin{theorem}[Hawking]\label{th:Hawking}
In a spacetime of sufficient differentiability, if there is a Cauchy hypersurface $\Sigma$ such that the trace $K$ of its second fundamental form satisfies
$K\geq b>0$ and the convergence condition (\ref{ConCon}) holds along the timelike geodesic congruence $v^\mu$ orthogonal to $\Sigma$
then all timelike geodesics are past incomplete.
\end{theorem}
Observe that $K$ is nothing but the expansion of the congruence $v^\mu$ at $\Sigma$. The idea therefore is that universes expanding everywhere will necessarily be past incomplete, hence this theorem is applicable to cosmology.

Since then many more singularity theorems have been proven, see e.g. \cite{HE,HP,Kri,S1,P5,SMilestone},
some of them are applicable to cosmological situations, some to star or galaxy collapse, and even others to the collision of gravitational waves, to cite some prominent situations.
Fortunately, all singularity theorems share a well-defined skeleton, the very same 
pattern. This is, succinctly, as follows \cite{S1,SMilestone}
\begin{theorem}[Pattern Singularity Theorem]
If a space-time of sufficient differentiability satisfies
\begin{enumerate}
\item a condition on the curvature
\item a causality condition
\item and an appropriate initial and/or boundary condition
\end{enumerate}
then there are null or time-like inextensible incomplete geodesics.
\end{theorem}
The paradigmatic case is the celebrated Hawking-Penrose  theorem \cite{HP}, which is still considered the singularity theorem {\em par excellence}. 
\begin{theorem}[Hawking and Penrose]\label{th:HP}
If the convergence condition (\ref{ConCon}) is satisfied, there are no closed future pointing timelike curves, a generic condition on the curvature holds and if there is one of the following:
\begin{itemize}
\item a compact spacelike (and achronal) hypersurface,
\item a closed trapped surface,
\item a point with re-converging light cone
\end{itemize}
then the space-time is causal geodesically incomplete.
\end{theorem}
The ``generic'' condition on the curvature is standard and amounts to having tidal effects some time somewhere for {\em every} possible causal curve. The achronality of the first option means that there are no two time-related points on the spacelike hypersurface, that is, it is a good ``instant of time''.

The ideas behind the assumptions in the pattern theorem are explained in some detail next.
\subsubsection{The curvature conditions}
These assumptions enforce the geodesic focusing via the Raychaudhuri equation. Many times they are
referred to as ``energy'' and/or ``generic'' condition, however as explained above the assumptions are of a geometric nature. A curvature condition is absolutely indispensable: no singularity theorem can be proven without some sort of such condition. The majority of the theorems, specially the stronger ones, use the condition (\ref{ConCon}) or averaged versions thereof, but sometimes some extra conditions on the full Riemann tensor are needed too, such as the generic condition in theorem \ref{th:HP}. 
\subsubsection{The causality condition}
 On the one hand, this prevents the possibility of traveling to one's own past, something not superfluous since G\"odel \cite{Go} showed in 1949 the existence of closed time-like lines in reasonable spacetimes.  More importantly, and basically, it ensures the existence of maximal geodesics between any two (causally related) events in appropriate subdomains of the spacetime.
For instance, a Cauchy hypersurface ensures the existence of such maximal geodesics in the entire spacetime. Otherwise, one restricts the analysis needed for the proof of the theorems to appropriate {\em globally hyperbolic} subregions of the spacetime. Observe that maximal geodesics cannot have focal points (caustics). For the general theory of causality see \cite{MSa}.
\subsubsection{The boundary/initial condition}
Recapitulating, {\em focusing of all causal geodesics} and thus existence of 
caustics and focal points follows from the curvature condition. But {\em the existence of maximizing geodesics}, hence necessarily without focal points, joining causally related events of the space-time also follows from the basic causality conditions. A contradiction starts to glimmer if all geodesics are 
complete. However, there is no such contradiction yet, because we have not enforced a finite upper bound for the proper time of selected families of time-like geodesics (and analogously for 
null geodesics). To get the final contradiction with geodesic completeness one needs to add the initial/boundary condition, which happens to be absolutely essential in the theorems. 
Later, I will discuss examples of physically reasonable singularity-free space-times for illustration.

\section{Critical evaluation of the theorems}
The singularity theorems are often quoted as one of the greatest theoretical
accomplishments in mathematical physics. They have tremendous physical consequences and cast serious doubts on the suitability of GR in very extreme situations. Nevertheless, the singularity theorems are often misquoted, and there are some ``myths'' and folklore about them that must be dispelled. As with other subtle matters, {\em the devil is in the detail}. In what follows I provide a selected lists of ideas concerning what the theorems say ---and do not say.

\subsection{(mis)-Interpretation of Penrose's theorem}
Sometimes the Penrose theorem is interpreted as definite proof that black holes form in gravitational collapse.
This is incorrect and the actual fact is much more subtle. The assumption of the existence of a trapped sphere \underline {does not} state anything about the formation of black holes, rather it uncovers (some of) what happens inside black holes once they are formed!
This follows from a theorem \cite{C,HE} stating that, in asymptotically flat spacetimes, no closed trapped surface can be seen from infinity, in other words, the trapped spheres are enclosed beyond the event horizon of the black hole, and thus they are already inside black holes {\em already formed}. 

Therefore, Penrose's theorem informs us of what happens {\em beyond the event horizon} of black holes. It is a result about the interior of black holes telling us that, if they form, inside them there will probably be closed trapped surfaces and, therefore, incompleteness of the spacetime will follow (classically and under certain conditions). This is not a minor thing, for hitherto this is one of the few solid informations we have been able to produce with some certainty about the {\em interior} of black holes. {\em Penrose's theorem uncovers part of the mysteries inside black holes}.

From the viewpoint of black-hole formation the question is quite another \cite{DZN}: to know if closed trapped surfaces form from innocuous initial data. There are very few rigourous results on this line, but some relevant advances were found in \cite{Chris1}, see also \cite{RT,KR}. 

\subsection{Geodesically complete spacetimes}
A way to understand the meaning of the singularity theorems is to turn things around and ask when geodesic completeness is feasible and what kind of models support it. Before giving the known general results on this matter, let me present several important illustrative examples:

\subsubsection{Einstein's static universe}
The Einstein universe metric ($\Lambda >0$)
$$
ds^2= -c^2 dt^2 + \frac{1}{\Lambda} (d\chi^2 +\sin^2\chi d\Omega^2)
$$
is a solution of (\ref{eq:EFE}) for a dust $T_{\mu\nu} = \varrho u_\mu u_\nu$ with $u_\mu =-\delta_\mu^t$, $8\pi G \varrho = 2c^4 \Lambda$ and $\chi\in(0,\pi)$ so that the 3-dimensional metric in parenthesis is the standard metric of a round 3-sphere.

This spacetime satisfies the convergence condition (\ref{ConCon}), is globally hyperbolic with compact spacelike slices ---any $t=$const. hypersurface is a compact Cauchy hypersurface---, contains points with reconverging light cones and, yet, it is geodesically complete.

The only assumption in the Hawking-Penrose theorem \ref{th:HP} that is not met is the {\em generic condition}, which can certainly be seen to fail for some specific timelike geodesics \cite{S1}.

\subsubsection{de Sitter spacetime}
\begin{figure}[!h]
\centering\includegraphics[width=12cm]{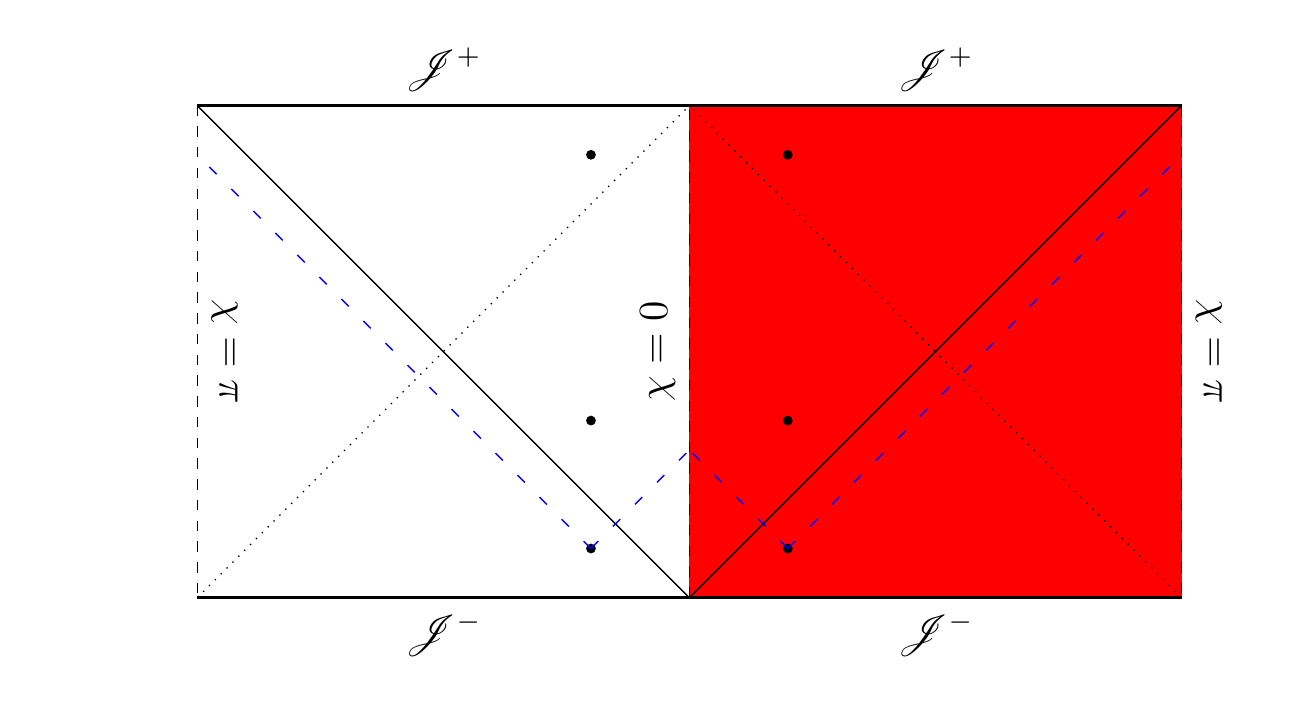}
\caption{A conformal diagram of de Sitter spacetime, null rays are at 45$^o$ and time goes upwards. The vertical lines $\chi=0$ and $\chi=\pi$ represent the world lines of the north and south poles of the spatial 3-sphere which shrinks for $t<0$ and then expands for $t>0$. Usually, only the red part of the diagram is represented \cite{HE}, however, our diagram provides a better representation for the compactness of the light cones emanating from some spheres (the trapped ones): one has to mentally identify the left and right $\chi=0,\pi$ lines producing a cylinder, and any sphere with constant values of $\chi$ and $t$ is then represented by two dots at the same height, as the three shown. The one below is a future trapped sphere, and its future has a boundary ---represented with the dashed lines--- which is {\em compact}. This spacetime escapes the conclusions of theorem \ref{genpen} because the Cauchy hypersurfaces are also compact, and then there is a faithful bijection between the dashed lines and any horizontal line. The sphere represented by the two dots in the middle is not trapped, and the one above is actually {\em past} trapped. \\
As an exercise for the interested reader, observe that de Sitter spacetime is fully homogeneous, and thus there is a future trapped sphere passing through every point. One can then choose {\em any} point in the diagram, and then try to identify the second point that conforms the {\em future} trapped sphere with the chosen point.}
\label{fig:dS}
\end{figure}

de Sitter spacetime is given by
$$
ds^2= -c^2 dt^2 + \lambda^2\cosh^2(ct/\lambda) (d\chi^2 +\sin^2\chi d\Omega^2), \hspace{1cm} \lambda^2=3/\Lambda
$$
where, as in the previous example, the $t=$const.\ 3-spheres are Cauchy hypersurfaces with the standard round metric, but now the spacetime is dynamic. This is a solution of (\ref{eq:EFE}) in vacuum ($T_{\mu\nu}=0$) so that (\ref{ConCon}) holds for null vectors. Round spheres with $t$ and $\chi$ constant are trapped for some values of the constants, namely when 
$$
\tan^2\chi > \frac{1}{\sinh^2 (ct/\lambda)}
$$
which, for any value of $t\neq 0$, is always feasible choosing $\chi$ appropriately. This is better understood in the conformal diagram of figure \ref{fig:dS}.

\vspace*{-7pt}

de Sitter spacetime is geodesically complete. It avoids theorem \ref{genpen} because the Cauchy hypersurfaces are compact (see figure \ref{fig:dS}), all other assumptions are satisfied. Theorems \ref{th:Hawking} and \ref{th:HP} do not apply either, despite having closed slices $t=$constant with positive expansion $K$, because the convergence condition does not hold along the timelike geodesics ---a $\Lambda$ effect.

\subsubsection{A singularity-free expanding perfect-fluid model}
The previous two examples are too symmetrical, and thus one may think that they are not representative enough. In this subsection a more subtle example is presented \cite{S,M}.

Take $(\mathbb{R}^4,g)$ in cylindrical coordinates $\{t,\rho,\varphi,z\}$ with line element
\begin{eqnarray}
ds^{2}&=&\cosh^4(at)\cosh^2(3a\rho)(-c^2dt^2+d\rho^2)\nonumber\\
&+&\frac{1}{9a^2}\cosh^4(at)\cosh^{-2/3}(3a\rho)\sinh^2(3a\rho)d\varphi^2 \nonumber\\
&+&\cosh^{-2}(at)\cosh^{-2/3}(3a\rho)dz^2, \label{sol}
\end{eqnarray}
where $a$ is a positive constant. This metric is a solution of (\ref{eq:EFE}) with $\Lambda=0$ for a perfect fluid energy-momentum tensor 
$$
T_{\mu\nu} = \varrho u_\mu u_\nu + p  \left(g_{\mu\nu} + u_\mu u_\nu \right)
$$
whose unit velocity one-form is
$$
u_{\mu}dx^\mu = -\cosh^2(at) \cosh(3a\rho)dt
$$
and energy density 
\be\label{density}
\frac{8\pi G}{c^4}\varrho = 15a^2\cosh^{-4}(at)\cosh^{-4}(3a\rho).
\ee
The equation of state 
$$
p=\frac{1}{3}\varrho
$$
is realistic for radiation dominated matter. Thus, all possible curvature conditions are satisfied, in particular the {\em strict} (\ref{ConCon}) (with strict inequality) holds for any $v^\mu$.

The spacetime possesses {\em cylindrical symmetry} with a regular axis at $\rho\rightarrow 0$ and is {\bf globally hyperbolic}, each $t=$const.\ slice is a non-compact global Cauchy hypersurface. Hence, the strongest causality requirements do hold. Observe that the energy density and pressure are regular everywhere: there are no curvature singularities and the spacetime is actually geodesically complete \cite{CFS}.

The trace of the 2nd fundamental forms (expansion) of the $t=$const.\ Cauchy hypersurfaces reads 
\be\label{eq:K}
K= \nabla_\mu u^\mu = 3a\frac{\sinh(at)}{\cosh^3(at)\cosh(3a\rho)} \, \, (>0 \, \, \mbox{for} \, \, t>0). 
\ee
which are positive for every $t>0$. The fluid is contracting for $t<0$, it rebounds at $t=0$, and then it expands for all $t>0$. The rebound is driven by the acceleration of the fluid
$$
a_{\mu}dx^\mu = 3a \tanh (3a\rho) d\rho
$$
arising due to the pressure gradients, which are forces opposing gravitational attraction.

At first sight it seems that this model satisfies all the assumptions of theorem \ref{th:Hawking}, as this is a realistic perfect fluid expanding {\em everywhere} for half of the history. Still all geodesics are complete. This can happen because $K$ in (\ref{eq:K}) is {\em not} bounded below by a positive constant, as $\lim_{\rho\rightarrow\infty} K =0$: a subtle property. Theorem \ref{genpen} does not apply because there are no closed trapped surfaces. Similarly, any of the choices for the boundary condition in theorem \ref{th:HP} does not hold. 

This model is not an exception, and there are entire families of such solutions \cite{RS} presenting reasonable physical properties. For further details, see \cite{S1,SRay}.

\subsubsection{General results for globally hyperbolic spacetimes}
Sometimes one can reverse the question and analyze the circumstances under which reasonable spacetimes can be geodesically complete. Thus, one asks: keeping the convergence condition (\ref{ConCon}), when \underline{globally hyperbolic} spacetimes can be geodesically complete?
There are a couple of results in this direction, the first for the stationary situation, the second for dynamical cases with the restriction that, at some instant of time, the {\em entire} spacetime expands (or contracts).

\paragraph{Stationary case}
In the \underline{stationary} case, it can be shown that if (\ref{ConCon}) holds then geodesic completeness requires \cite{GaH}
$$
\frac{R_{\mu\nu}\xi^{\mu}\xi^{\nu}}{\xi^{\mu}\xi_{\mu}}\sim k/\bar\rho^2
$$ 
for some constant $k$, where $\vec\xi$ is the timelike Killing vector field while $\bar\rho$ is an appropriate spatial distance between any two events. 

\paragraph{Dynamical cases}
In the \underline{dynamical} and open case, it can be proven that if (\ref{ConCon}) holds and the expansion $K$ of a {\em non-compact}\footnote{This is no real restriction because, if the Cauchy hypersurface is compact, there are stronger singularity theorems that apply \cite{HE}: closed expanding non-singular models necessarily require the violation of (\ref{ConCon}), see \cite{S1}.} Cauchy hypersurface $\Sigma$ is everywhere positive then at least one of the following three quantities must be non-positive \cite{SRay,S3} 
\begin{itemize}
\item the cosmological constant $ \Lambda $
\item the \underline {averaged} energy density on $ \Sigma $
\item minus the \underline {averaged} scalar curvature of $ \Sigma $
\end {itemize}
Here the average of any scalar quantity $Q$ is defined as
$$
\lim_{S\rightarrow \Sigma} \frac{\int_S Q}{\mbox{Vol} (S)}
$$
where the limit must be taken with care, as there arise some mathematical subtleties \cite{S3}. A similar theorem can be proven if $K$ is negative.  

In both the stationary and dynamical cases the conclusion can be stated as {\em ``the energy density must fall off quickly enough in spatial directions''}. This is a very satisfactory result, that can be read as: regular cosmological models are not viable if we believe that the Universe has a more or less homogeneous distribution of matter everywhere.

Situations where the expansion $K$ changes sign over the non-compact $\Sigma$ have not been considered hitherto, to my knowledge.

\subsubsection{Regular black holes}
There are plentiful models for regular black holes in the literature, one may consult the recent analysis in \cite{Maeda} to get an idea of their properties and the different possibilities. For the spherically symmetric case, most of them share similar properties from the viewpoint we are interested in here, thus I will use a particularly interesting model presented in \cite{MMS} which keeps the entire exterior region to be the same as in the Schwarzachild vacuum (with $\Lambda =0$) solution . 

\begin{figure}[!h]
\centering\includegraphics[width=8cm]{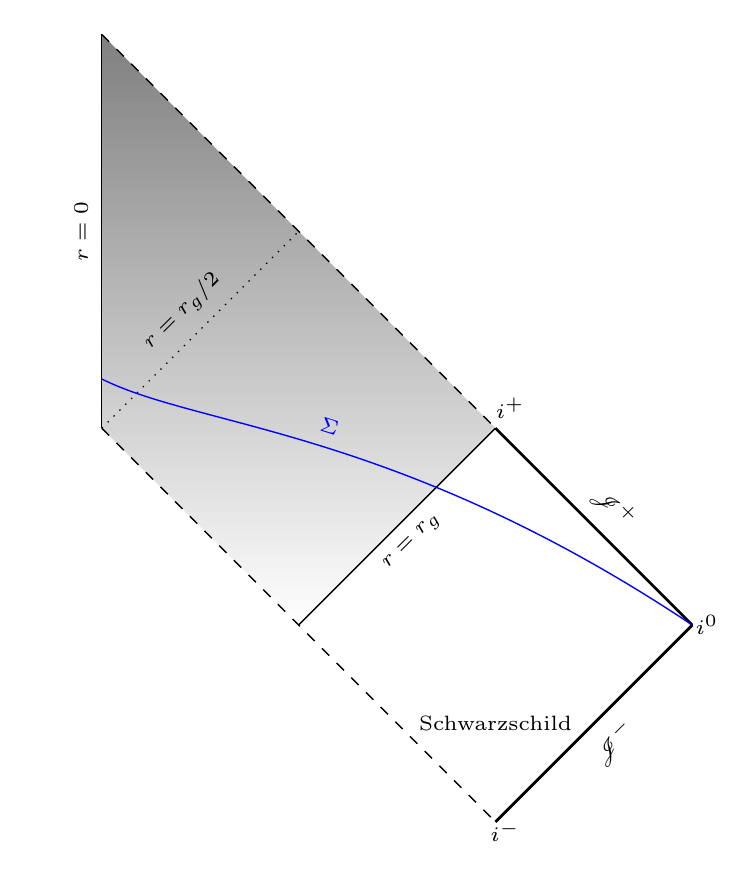}
\caption{A conformal diagram of the spacetime with metric \ref{eq:BHregular}, null rays are at 45$^o$ and future goes upwards. Every point represents a round 2-sphere except for the line $r=0$, which is a regular centre. The non-shadowed region represents exactly the asymptotically flat Schwarzschild vacuum solution, with its standard asymptotic structure. At $r=r_g =2GM/c^2$ the metric is matched to another mass function such that the entire spacetime is free from curvature singularities. The region between $r=r_g$ and $r=r_g/2$ contains trapped round spheres. The hypersurface $\Sigma$ shown in blue is a non-compact Cauchy hypersurface. Given that the null convergence condition holds, theorem \ref{genpen} applies and some null geodesics cannot be complete. This can be easily seen to happen for null (and also timelike) geodesics reaching the dashed lines, to the future and to the past. Still, the spacetime can be {\em regularly extended} beyond these dashed lines, and the extension can be so chosen such that the whole spacetime is geodesically complete. In the extended spacetime Penrose's theorem cannot apply, and as the null (\ref{ConCon}) can be easily kept, the only possibility is that $\Sigma$ fails to be, in the extended and complete spacetimes, a Cauchy hypersurface. See fig.\ref{fig:BHextended}.}
\label{fig:BHregular}
\end{figure}


The line-element reads, in standard advanced coordinates
\be\label{eq:BHregular}
ds^2= -e^{4\beta(r)} \left(1-\frac{2\mu(r)}{r}\right) dv^2 +2dvdr +r^2 d\Omega^2
\ee
where the functions $\beta$ and $\mu$ depend only on $r$ and are given, explicitly, by
\bean
2\mu (r) &=& r_g \Theta(r-r_g) + \frac{r^3}{r_g^2} \left(10-15\frac{r}{r_g} +6\frac{r^2}{r_g^2} \right)\Theta(r_g-r)\\
\beta (r) &=& \frac{5}{12}\left(1+3\frac{r}{r_g}\right)\left(\frac{r^3}{r_g^3} -1\right)\Theta(r_g -r) 
\eean
where $\Theta (x)$ is the step Heaviside function and 
$$
r_g := \frac{2GM}{c^2}
$$
is the gravitational radius for a mass $M$. This spacetime coincides exactly with the Schwarzschild vacuum solution (with $\beta =0$ and $2\mu = r_g$) for all $r\geq r_g$. For $r< r_g$ the spacetime acquires a matter content (in GR) and one can prove that the weak energy condition \cite{HE,Maeda} holds, in particular, the {\em null} convergence condition (\ref{ConCon}) is satisfied. The null hypersurface $r=r_g$ is the event horizon and the region $r<r_g$ contains another value, $r=r_g/2$, such that $2\mu(r_g/2)=r_g/2$, with $2\mu(r)>r$  in between. Hence, there are trapped round spheres for all $r\in(r_g/2,r_g)$, see figure \ref{fig:BHregular}.
There are no curvature singularities anywhere, the value $r=0$ is a regular centre of symmetry. There also exist non-compact Cauchy hypersurfaces. Therefore, according to theorem \ref{genpen} some null geodesics must be incomplete. It is easily seen that this is the case for the radial null geodesics not in the $v=$ const.\ hypersurfaces, because they reach $r=r_g$ towards the past, or $r=r_g/2$ to the future, with finite affine parameter. However, the metric can be regularly extended through these values of $r$, and in particular {\em extensions} can be produced which are geodesically complete \cite{MMS}, as shown in fig.\ref{fig:BHextended}. These extensions avoid the singularity theorems as explained in the figure.

This leads me to the important problem of spacetime extensions.

\subsection{The problem of extensions}
The geodesic incompleteness proven by the singularity theorems may simply indicate incompleteness of the manifold itself, such as excising regular points, with no relation whatsoever to curvature, conical, or similar singularities. This poses an important physical problem: how to deal with extensible space-times? 
The answer may seem simple: just extend them until you cannot extend it any more. However, this is not so simple at all.

\begin{figure}[!h]
\centering\includegraphics[width=10cm]{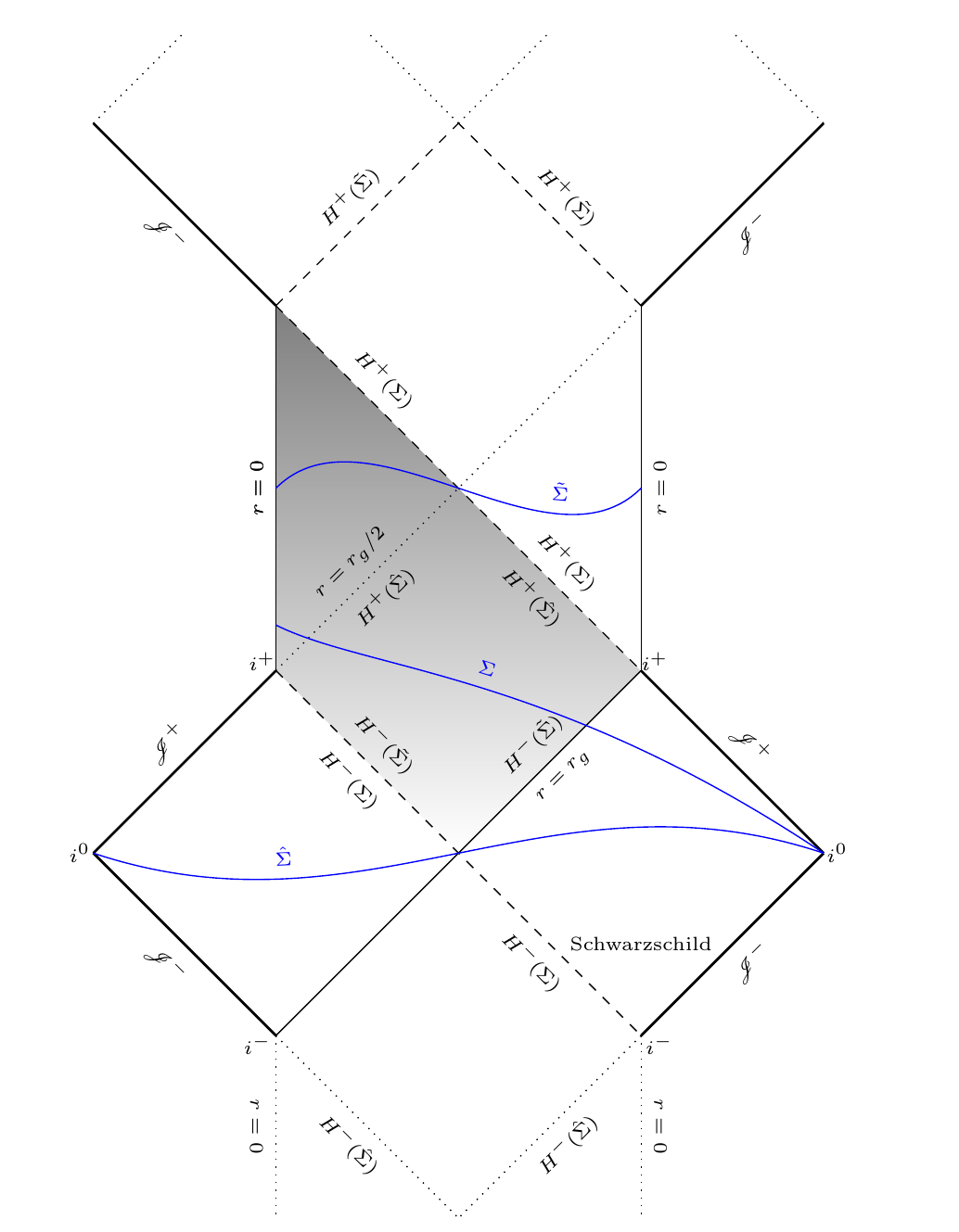}
\caption{A conformal diagram of the spacetime with metric \ref{eq:BHregular} extended beyond the dashed lines of the previous figure \ref{fig:BHregular}, null rays are at 45$^o$ and future goes upwards. Every point represents a round 2-sphere except for the lines $r=0$, which are regular centres. The extension can be performed in an infinite number of ways; here I have chosen a completely regular one that can be easily built by gluing together $\infty$ different copies of the original spacetime, see \cite{MMS} for details. The diagram continues upwards and downwards indefinitely as indicated by the dotted lines. Three inequivalent slices are shown, denoted by $\Sigma,\tilde\Sigma$ and $\hat\Sigma$ in blue. None of them is a true Cauchy hypersurface, they are only partial ones \cite{HE,S1,HP}. The corresponding past and future Cauchy horizons are shown. Thus, theorem \ref{genpen} is not applicable; the more general theorem \ref{th:HP} does not apply either because the {\em timelike} convergence condition (\ref{ConCon}) fails. These type of regular black holes always exhibit {\em change of topology}, meaning that there are slices (spacelike and achronal hypersurfaces with no edge) with different topologies. In our case, the orginal $\Sigma$ has topology $\mathbb{R}^3$, while $\hat\Sigma$ and $\tilde\Sigma$ are topologically $\mathbb{R}\times \mathbb{S}^2$ and $\mathbb{S}^3$, respectively.}
\label{fig:BHextended}
\end{figure}

Extensions are highly non-unique. As an obvious example, observe that the metric (\ref{eq:BHregular}) is actually an extension of the exterior region $r>r_g$ of Schwarzschild vacuum solution, but this is different from the standard analytical one found in the textbooks \cite{Wald,HE}. In fact,  physically meaningful extensions are far from obvious and, furthermore, not even analytical extensions are unique, nor feasible, in general, let alone smooth extensions. For explicit examples with a discussion see \cite{FMS}. This problem is especially relevant at {\em Cauchy horizons}, see \cite{HE,S1,Wald,P5} for definitions and figure \ref{fig:BHextended} for several illustrative examples. The regularity class of the extension is crucial here, see for recent mathematical results \cite{Sb,DL} and references therein.

For a given extensible space-time, there are generically inequivalent extensions leading to 
\begin{enumerate}
\item new extensible space-times. This is for instance the case of (\ref{eq:BHregular}), which as shown in fig.\ref{fig:BHregular} is a non-typical extension of the exterior vacuum Schwarzschild spacetime.
\item to geodesically complete and inextensible spacetimes, such as the extension of the previous extension shown in figure \ref{fig:BHextended}.
\item geodesically incomplete, inextensibe space-times, such as the standard Kruskal extension \cite{HE,Wald}.
\end{enumerate}
It might seem advisable that one should choose (ii), but this is not usually the case: if the singularity-free extension violates a physical condition, such as causality or energy positivity, then 
other extensions might be preferred. Keep in mind, however, that this is not always the chosen way to proceed: the standard maximal analytical extension of Kerr spacetime \cite{HE,GP} leads to causality violation, as an outstanding example. 

In summary, which physical reasons are to be used to discriminate between inequivalent extensions? And, are we always choosing the extensions in a  coherent form?

\section{The character of the singularity}
The singularity theorems give very little information about the nature of singularities,
they merely state that for some unspecified reason some particle or light ray ends suddenly (future singularity), or appears {\em ex nihilo} (past one). Nevertheless, Belinskii, Khalatnikov and Lifschitz (BKL) \cite{BKL,BKL1} argued, based on the field equations (\ref{eq:EFE}), that for generic dynamical situations one can concentrate on dominant terms --given by the time derivatives of the variables-- to get a faithful idea of the character of the singularity. 

\begin{figure}[!h]
\centering\includegraphics[width=10cm]{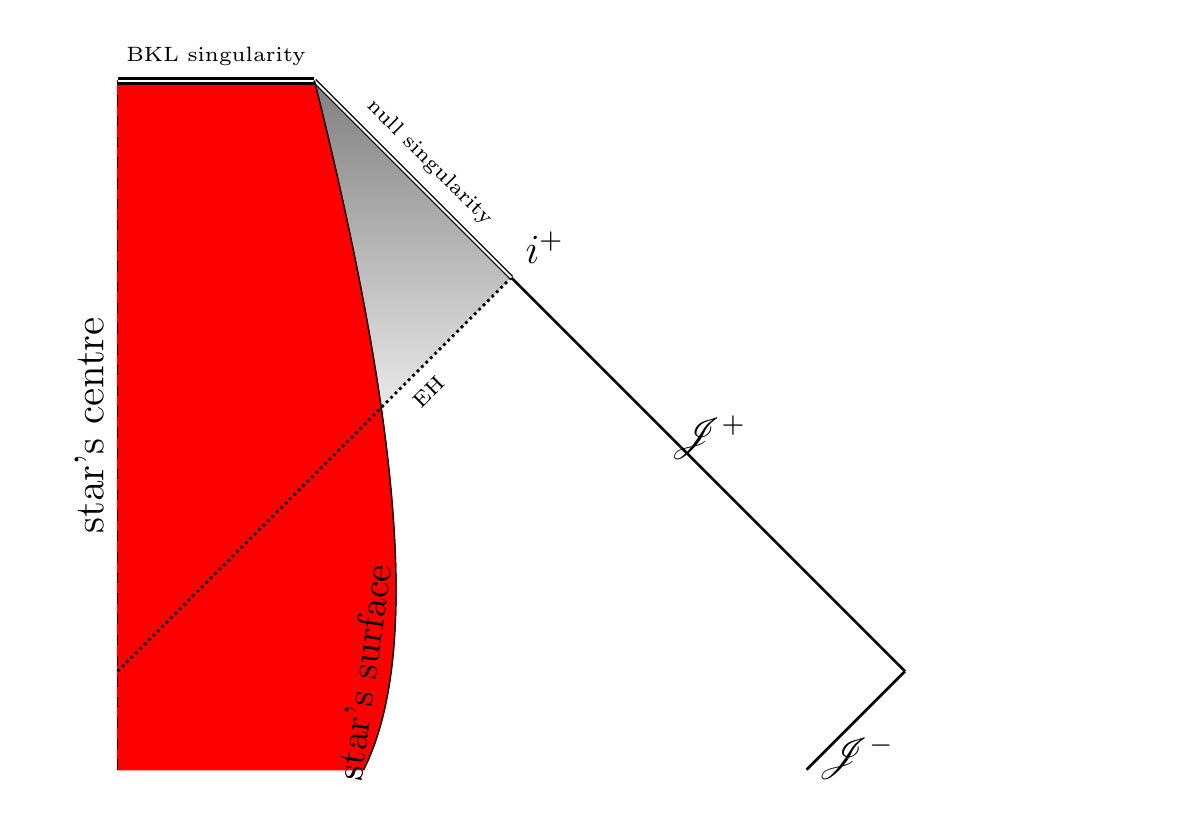}
\caption{A conformal diagram of an approximately spherical spacetime representing the gravitational collapse of a self-gravitating body (the red region) to form a black hole. As usual, radial null rays are at 45$^o$ and future goes upwards. Every point represents a 2-sphere except for the line marked as the star's centre. The event horizon of the black hole is represented by EH, the shadowed zone is the vacuum portion of the black hole region, while the asymptotically flat  exterior has the standard future null ($\scri^+$) and timelike ($i^+$) infinities. From rigourous mathematical results \cite{DL} one knows that, {\em generically}, there is going to be a {\em null} piece of the spacetime boundary to the future of $i^+$. This could be either a mildly regular Cauchy horizon or a weak singularity, and is represented here as the ``null singularity''. If the entire singularity were null, the BKL picture would be false. However, there are recent results in spherical symmetry \cite{V} proving that this cannot be the case: there must be another {\em non-null} piece of the singularity. This is represented here by the ``BKL singularity'', suggesting that perhaps the BKL conjecture should be applied only to this portion of the singularity. Whether or not this is the right picture of a realistic gravitational collapse is yet unknown.}
\label{fig:BKL+null}
\end{figure}

The BKL conclusion can be summarized as {\em singularities are spacelike, local, oscillatory, and ‘matter does not matter’} \cite{SMilestone}. By spacelike is meant that this occurs at an `instant of time $t$', where $t$ is the time variable used in an appropriate synchronous (or Gaussian) coordinate system. The term local refers to neglecting the spatial derivatives, so that the effective dynamics at each point decouples from that of neighbouring points and is well described by the evolution of a spatially homogeneous (Bianchi) model, which happens to be oscillatory. Finally, matter can be safely ignored because the anisotropic terms diverge faster than the matter terms and thus the approach to the singularity is well described by (\ref{eq:EFE}) with vanishing $T_{\mu\nu}$ and $\Lambda =0$. All this is usually known as the BKL conjecture, which is supported by mathematical results \cite{Ringstrom} as well as numerical ones \cite{G1,Ber} ---sea also references therein.

Though the BKL picture is consistent with the field equations there are other possibilities on the market, the main competitor is the idea of {\em null singularities} advocated in \cite{PI}, which is actually based on the instability of Cauchy horizons. The Cauchy horizons of a Reissner-Nordstr\"om, or a Kerr, or a regular black hole --as those shown in figure \ref{fig:BHextended}-- are \underline{unstable} null hypersurfaces in the sense that small perturbations blow up there. These perturbations should turn the horizon into a singularity but without losing its null character. Invoking the black-hole  uniqueness theorems, this picture should also lead to a good description of the singularity inside a black hole. The existence of a null singularity is supported, too, by theoretical, numerical and mathematical arguments \cite{OF,BS,D,DL} . 

Thus, there is some tension between the BKL and Null pictures. Recent mathematical results show that, generically, there will always be at least a portion of a null singularity for realistic gravitational collapse starting from good initial data, see figure \ref{fig:BKL+null} to see the placement of this null portion. However, this null part does not have to cover the entire singularity and, actually, there are recent results proving that (in spherical symmetry) such a null portion {\em cannot be} the entire singularity \cite{V}. Hence, there must be  a non-null part of the singularity. This portion may well be appropriately described by the BKL conjecture, see figure \ref{fig:BKL+null}. 

If this combined picture is correct, then collapsing matter within the star should eventually encounter a BKL singularity. On the other hand, an observer who enters into the black hole but keeps permanently outside the star should meet a null singularity.

\section{Conclusions in brief}
Sometimes one reads or hears that ``singularities are consubstantial to GR'' and such a claim is supposedly  based on the singularity theorems. Well, nothing of the kind: almost all gravitational systems (planets, stars, planetary systems, clusters, galaxies, pulsars, binary systems, etc.) are regular and accurately described with GR and its (post-Newtonian) limits. The only possible exceptions are the very early Universe and the --somehow mysterious-- black holes' deep interiors. The incompleteness predicted by the theorems in these two exceptions are certainly \underline{a distinctive feature of GR}. Therefore, the singularity theorems provide supporting evidence for the need of (quantum? \cite{Boj,Wall}) corrections to GR \underline{well inside black holes} and probably at the initial stages of our Universe \cite{BV2}.

There also remain important problems to be clarified. In particular, the question of extensions, when and how they must be performed, and under which criteria. Also, the type and character of the singularity is still uncertain. 

The most powerful theorems, such as the Hawking-Penrose theorem \ref{th:HP}, have little application as they do not state where and when is the singularity, nor its strength or size. Thus, the past big-bang singularity will be enough to comply with the theorem and no other prediction would follow.

Finally, as an aside remark, to this day there is no theorem available that can ``predict'' the singularity in the (analitycally maximally extended) Kerr black hole despite very interesting and clever recent attempts \cite{L,Mi,Mi0}. Will there ever be one?

\enlargethispage{20pt}




\competing{The author declares that he has no competing interests.}

\funding{Work supported under Grants No. FIS2017-85076-P (Spanish MINECO/AEI/FEDER, EU) and No. IT956-16 (Basque Government).}

\ack{I am grateful to the editors for inviting me to write this contribution.}



\end{document}